\documentclass[12pt]{article}

\usepackage{scicite}

\usepackage{times}
\usepackage{amsmath}
\usepackage{graphicx}

\newenvironment{sciabstract}{%
\begin{quote} \bf}
{\end{quote}}

\title{The Fourier series solution of three-body problem}

\author
{Siwei Luo$^{1\ast}$ 
\\
\normalsize{$^{1}$Jiangxi Normal University,}\\
\normalsize{No.99 Ziyang Avenue, High Technology Development Zone, Nanchang, Jiangxi, 330022, China}\\
\\
\normalsize{$^\ast$E-mail:  luosiwei@jxnu.edu.cn}
}

\date{}

\begin{document}

\baselineskip24pt


\maketitle


\begin{sciabstract}
The three-body problem is essentially to solve three curves that satisfy Newton’s equations. Given initial conditions found in numerical simulation, this paper introduces the Antikythera algorithm that solves three-body problem Fourier series solution via the Runge-Kutta method and Fourier transform. The Lagrange, BHH, figure-8, and IA100 solutions are reported along with their spectrum and parameter values.
\end{sciabstract}

\section*{Introduction}
The three-body problem that calculates trajectories of three particles under universal gravitation given their masses and initial conditions was proposed by Newton in 1687\cite{Newton}. In a system composed of 3 particles, the mass of $i$th particle is $m_i$, and the resultant force of other particles on it is\cite{Newton}:
\begin{equation}
    m_i\boldsymbol{\ddot{r}_i} = \sum_{j=1,j\neq i}^{3} G m_i m_j \frac{\boldsymbol{r_j}-\boldsymbol{r_i}}{|\boldsymbol{r_j}-\boldsymbol{r_i}|^3}  
    \centering
\end{equation}
where, geometrically, the trajectory curve of the $i$th particle in planar Newtonian three-body problem is a mapping $\boldsymbol{r}_i:{R} \rightarrow {R^{2}}$. Without loss of generality, set the gravitational constant G to be 1. Thus, the governing equations of three-body system is: 
\begin{equation}
\centering
\begin{cases}
    \ddot{\boldsymbol{r}}_1 = \sum\limits_{j=1,j\neq i}^{3} m_j \frac{\boldsymbol{r}_j-\boldsymbol{r}_1}{|\boldsymbol{r}_j-\boldsymbol{r}_1|^3}  \\
    \vspace{0.01cm} \\ 
    \ddot{\boldsymbol{r}}_2 = \sum\limits_{j=1,j\neq 2}^{3} m_j \frac{\boldsymbol{r}_j-\boldsymbol{r}_2}{|\boldsymbol{r}_j-\boldsymbol{r}_2|^3}  \\
    \vspace{0.01cm} \\ 
    \ddot{\boldsymbol{r}}_3 = \sum\limits_{j=1,j\neq 3}^{3} m_j \frac{\boldsymbol{r}_j-\boldsymbol{r}_3}{|\boldsymbol{r}_j-\boldsymbol{r}_3|^3}  
\end{cases}
\end{equation}
In 2-dimensional Euclidean coordinates, the differential equations along each axis for the $ith$ particle are:
\begin{equation}  
\begin{cases}
    {\ddot{x_i}} = \sum\limits_{j=1,j\neq i}^{3} m_j \frac{{x_j-x_i}}{((x_j-x_i)^2+(y_j-y_i)^2)^{\frac{3}{2}}} \\
    \vspace{0.01cm} \\
    {\ddot{y_i}} = \sum\limits_{j=1,j\neq i}^{3} m_j \frac{{y_j-y_i}}{((x_j-x_i)^2+(y_j-y_i)^2)^{\frac{3}{2}}} 
\end{cases}
\end{equation}
Solving the three-body problem means to calculate three curves $\{\boldsymbol{r}_1,\boldsymbol{r}_2,\boldsymbol{r}_3\}$ that satisfy Newton's equations, namely,  
\begin{equation}
\begin{cases}
\boldsymbol{r}_1(t) = (x_1(t),y_1(t)) \\ 
\vspace{0.01cm} \\
\boldsymbol{r}_2(t) = (x_2(t),y_2(t)) \\
\vspace{0.01cm} \\
\boldsymbol{r}_3(t) = (x_3(t),y_3(t)) \\
\end{cases}
\end{equation}
The three-body problem statement can be naturally extended to the n-body problem in 3-dimensional space. 

Although the three-body problem has been studied from many different perspectives, the solutions of three-body problem are always the most important and fundamental aspect. The Euler-Lagrange family solutions indicate the Newton's equations has analytical solution with collinear or equilateral configurations. The Broucke-Hadjidemetriou-Hénon(BHH) family solution is considered to be the closest to the orbits of three-body systems observed in real astrophysics. Its characteristics is that one orbit is close to a circle, and the other two are rose curves\cite{Broucke, Hadjidemetriou, Hénon}. For the equal-mass-zero-angular-momentum family solution, the entire pattern of the three orbits exhibits centrosymmetry with respect to the center-of-mass\cite{Li}. The figure-8 solution is found by Moore in 1993. Three particles with the identical mass are moving along the same trajectory of the shape of figure eight and the total angular momentum of the system is zero. The concise and elegant orbit is a fascinating study in dynamical system and astrophysics \cite{Moore, Chenciner}. 

In this paper, we introduce the Antikythera algorithm to calculate the Fourier series solutions given their initial conditions and present the results of Lagrange, BHH, figure-8, IA100 solutions. 

\section*{The Antikythera algorithm for the three-body problem}

The existence of convergent infinite series solutions for the three-body problem was studied by Karl Frithiof Sundman\cite{Sundman}. The Fourier series is complete to outline curves in space. For the periodic solutions of the three-body problem, the Fourier series parameter equations of the curves are suitable for describing orbits of the three-body problem. Compared with numerical solutions, Fourier series solutions are differentiable, they are more important for proving the Newton's equations and more informative for evolution of dynamical system. 

The Antikythera machine, found in a sinked ship, is the bronze mechanical computer trace the motion of celestial bodies. Following the same spirit we develop the Antikythera algorithm for computing the Fourier series solution of three-body problem. 

The position and velocity of each particle moving in the 2-dimensional plane is a 4-dimensional vector in the phase space. Therefore, the state of 3 particles is a 12-dimensional vector $\phi$ in the phase space. The $\phi_{k+1}$ state in phase space can be calculated based on the $\phi_{k}$ state. Given the initial state of three particles in phase space, the trace of the evolution of three-body system will be calculate via the dynamical system. 

The numerical approach like Runge-Kutta method transfer the original Newton's differential equations to a dynamical system. Given the initial conditions of three particles, the sequence of discrete points in phase space is calculated with numerical approach. After calculating the motion of three-body system in certain time interval, the sequence of positions and velocities information of each particle are obtained. 

The Fourier transform calculates the inner product of $x_i$ and trigonometric functions and the corresponding frequencies in $x_i$ and $y_i$ exhibits peaks in Fourier analysis spectrum\cite{Fourier}. The reconstructed parametric curve equation of the $i$th particle is written as: 
\begin{equation*}
\begin{split}
\boldsymbol{r}_{i}(t) = \left \{
\begin{array}{ll}
    x(t) = x_0 + a_0 sin(2\pi\omega_0 t+\phi_0) + a_1 sin(2\pi\omega_1 t+\phi_1) + ... \\               
    y(t) = y_0 + b_0 sin(2\pi\alpha_0 t+\psi_0) + b_1 sin(2\pi\alpha_1 t+\psi_1) + ... 
\end{array}
\right.
\end{split}
\centering
\label{eqn:1} \tag{2} 
\end{equation*}
The goal is to solve parameters set $\{x_0,\omega_k,\phi_k\,y_0,\alpha_k,\psi_k\}$ in the curve parametric equations. Parseval's theorem, a generalization of the Pythagorean theorem in Hilbert spaces, states the sum of the square of a function is equal to the sum of the square of its Fourier transform\cite{Parseval}. 

Both the amplitudes and frequencies of the corresponding trigonometric functions are calculated by Fourier transform. Construct the $x(t)$ and $y(t)$ functions with obtained amplitudes and frequencies and fit constants $x_0$ and $y_0$ and phase parameters $\phi_k$ and $\psi_k$, then all parameters of parametric curve equations are solved.

\begin{figure}[h!]
    \centering
    \includegraphics[width=0.8\textwidth]{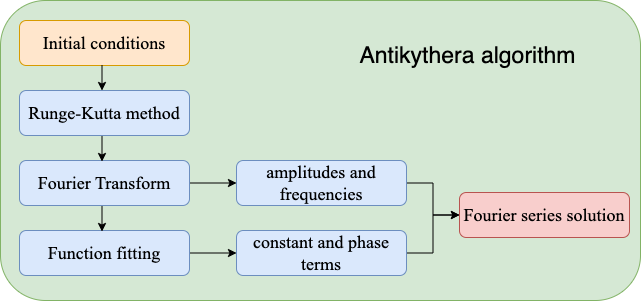} 
    \caption{The Antikythera algorithm for the three-body problem.}
    \label{fig:algorithm}
\end{figure}

In summary, the flowchart of the Antikythera algorithm is shown in Figure.\ref{fig:algorithm}, which consists of three main steps: 1) compute the numerical solution of three-body system with numerical method such as Runge-Kutta method; 2) conduct Fourier analysis on the numerical solution to obtain the frequencies and amplitudes; 3) fit constant terms $x_0$ and $y_0$ and phase terms $\phi_k$ and $\psi_k$ with obtained frequencies and amplitudes. Then write down the Fourier series solution of three-body problem, which is differentiable, with all calculated parameters. In following sections, the results from Lagrange, BHH, and equal-mass-zero-angular-momentum families are analyzed. 

\section*{The Lagrange solution}

In the Lagrangian solution, three particles are located at three vertices of an equilateral triangle. During the movement, the orbits of the three particles are three ellipses and the length and direction of the equilateral formed by the three particles changes but always remains equilateral, i.e., $||\boldsymbol{r}_1 - \boldsymbol{r}_2|| = ||\boldsymbol{r}_1-\boldsymbol{r}_3|| = ||\boldsymbol{r}_2-\boldsymbol{r}_3||$. Because the total force exerted on each particle from the other two always pass through the center-of-mass, the Lagrange three-body problem can be solved with similar approach to the two-body problem\cite{Lagrange}. 

An example of Lagrange solution is that the masses of three particles are $m_1 = 0.03, m_2 = 0.02, m_3 = 0.04$ and the initial conditions are: \\
$\boldsymbol{r}_1(0) = (0.19245008972987526, 1.0);$ \\
$\boldsymbol{r}_2(0) = (1.058475493514314,-0.49999999999999994);$ \\
$\boldsymbol{r}_3(0)=(-0.6735753140545633,-0.49999999999999994);$ \\
$\dot{\boldsymbol{r}}_1(0)=(-0.07896444077714955, 0.015196713713031851);$ \\
$\dot{\boldsymbol{r}}_2(0) = (0.03948222038857477, 0.08358192542167518);$ \\
$\dot{\boldsymbol{r}}_3(0)=(0.03948222038857477,-0.05318849799561147)$. \\
\begin{figure}[h!]
    \centering
    \includegraphics[width=0.9\textwidth]{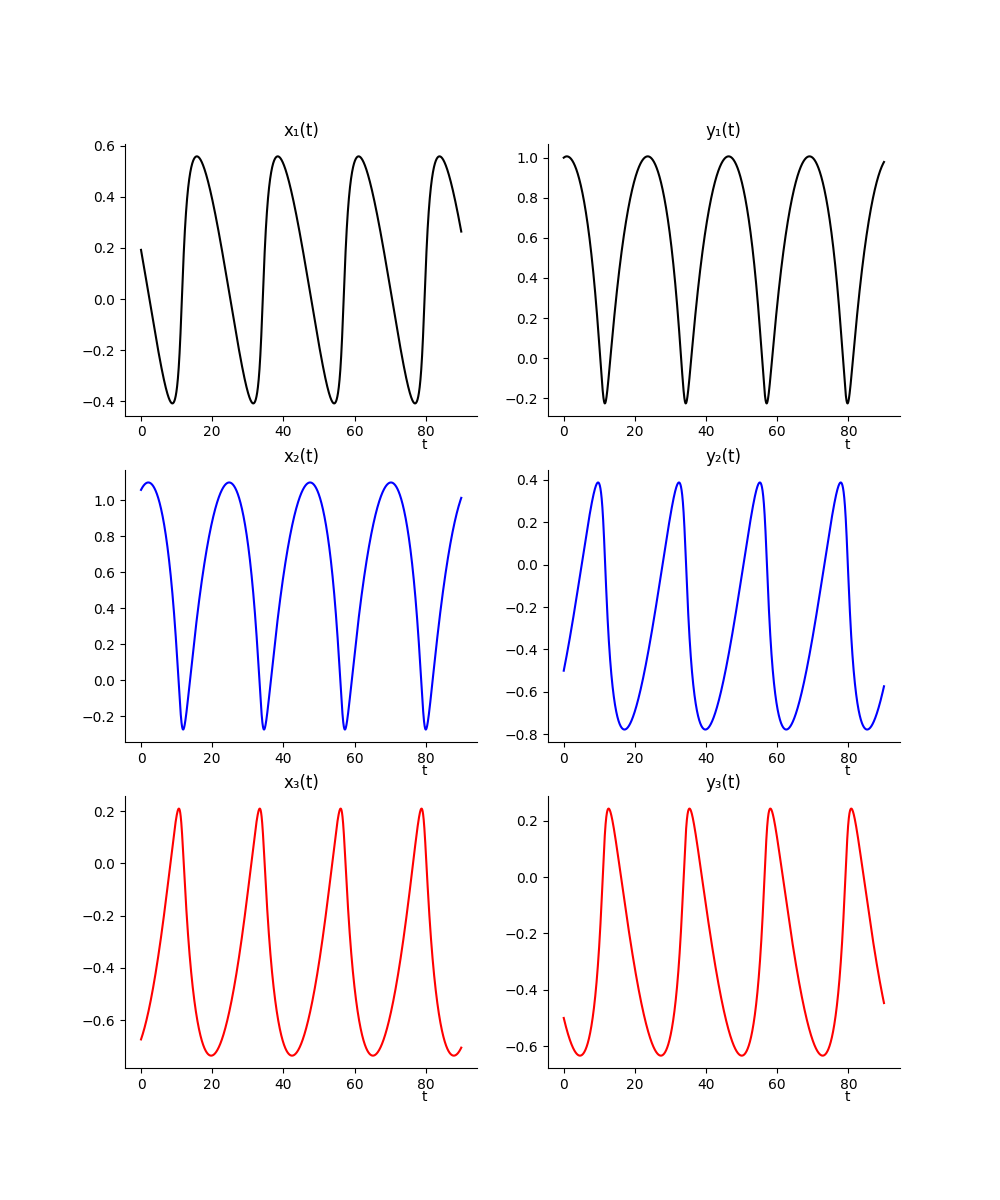} 
    \caption{The $x(t)$ and $y(t)$ of three particles' orbits from numerical solution.}
    \label{fig:my_label}
\end{figure}
Calculated by the Antikythera algorithm, the spectrum of orbits of three particles is shown in the Figure.\ref{Lagrange_spectrum}. 
\begin{figure}[h!]
    \centering
    \includegraphics[width=0.9\textwidth]{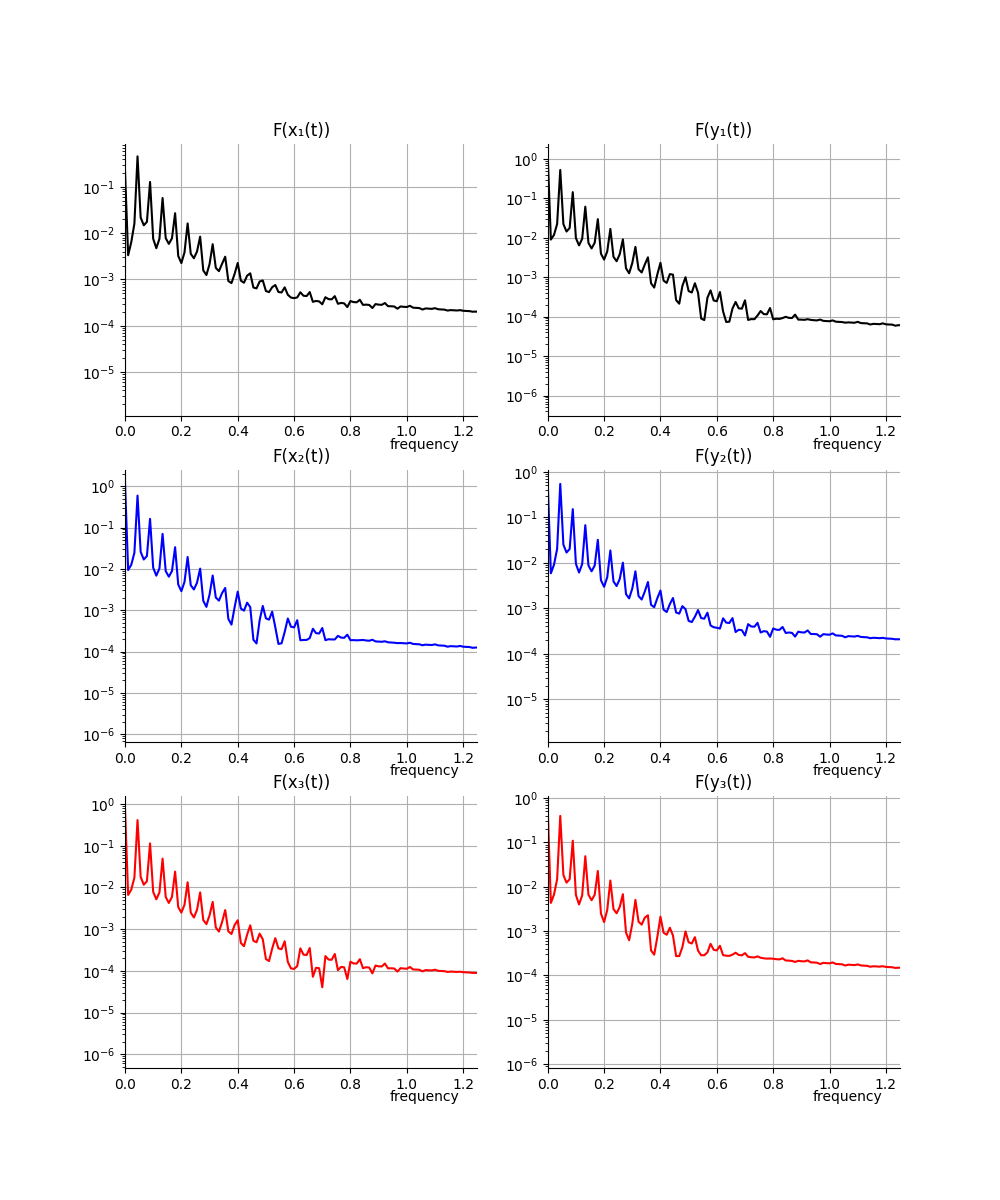} 
    \caption{The Fourier analysis of three particles' orbits.}
    \label{Lagrange_spectrum}
\end{figure}
From the empirical experimental result, the base frequency is 0.044444444 and the other frequencies are multiple of base frequency.

\begin{figure}[h!]
    \centering
    \includegraphics[width=0.9\textwidth]{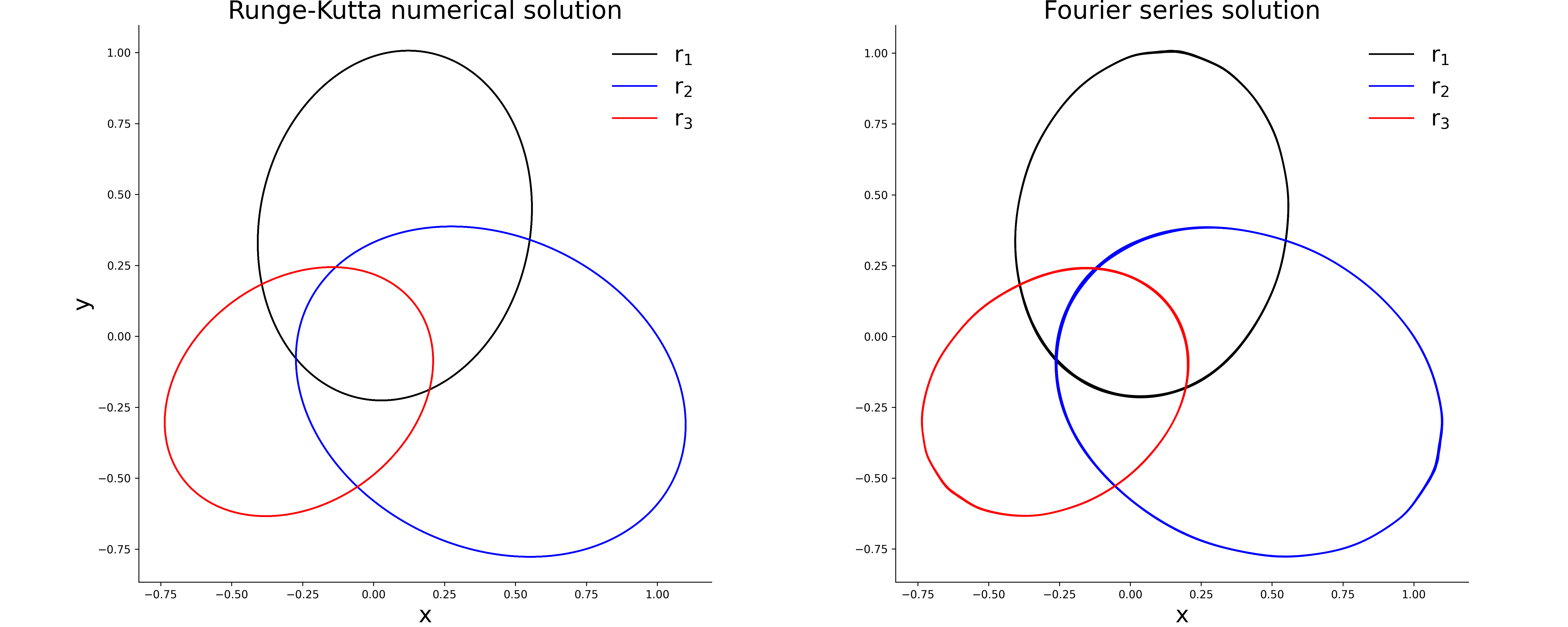} 
    \caption{The Runge-Kutta numerical solution(left) and the Fourier series solution(right) of the Lagrange solution.}
    \label{Lagrange_RungeKutta_Fourier}
\end{figure}
The Figure.\ref{Lagrange_RungeKutta_Fourier} shows the comparison between Runge-Kutta numerical solution and Fourier series solution. The Fourier analysis of Lagrange solution reveals the Fourier series form of ellipse. 

\begin{equation*}
\begin{split}
\boldsymbol{r}_{1}(t)= \left \{
\begin{array}{ll}
    x_{1}(t) \approx 0.111456457+0.458215129*sin(0.044444444* 2\pi t+2.794701043327) \\
    \hspace{1.5cm} + 0.128067684*sin(0.088888888*2\pi t-0.47929680311) + ... \\       
    y_{1}(t) \approx 0.5811966452+
    0.524970110*sin(0.044444444*2\pi t+1.27664477069) \\
    \hspace{1.5cm} + 
    0.143567818*sin(0.088888888*2\pi t-1.9979199642) + ... 
\end{array}
\right.
\end{split}
\centering
\label{Lagrange_r1}
\end{equation*}

\begin{equation*}
\begin{split}
\boldsymbol{r}_{2}(t)= \left \{
\begin{array}{ll}
    x_{2}(t) \approx 0.6152949512+0.524970110*sin(0.044444444* 2\pi t+1.05406007798) \\
    \hspace{1.5cm}+ 0.162118935*sin(0.088888888*2\pi t-2.2252346931) + ... \\       
    y_{2}(t) \approx -0.290158659+
    0.539333087*sin(0.044444444*2\pi t-0.6283207579) \\
    \hspace{1.5cm}+ 0.150404024*sin(0.088888888*2\pi t+2.38642908793) + ... 
\end{array}
\right.
\end{split}
\centering
\label{Lagrange_r2}
\end{equation*}

\begin{equation*}
\begin{split}
\boldsymbol{r}_{3}(t)= \left \{
\begin{array}{ll}
    x_{3}(t) \approx -0.391239535+0.413994573*sin(0.044444444* 2\pi t-1.1293050516) \\
    \hspace{1.5cm}+ 0.114373008*sin(0.088888888*2\pi t+1.88982558405) + ... \\       
    y_{3}(t) \approx -0.290818163+
    0.397609775*sin(0.044444444*2\pi t-2.5603602521) \\
    \hspace{1.5cm}+ 0.109689082*sin(0.088888888*2\pi t+0.43735828509) + ... 
\end{array}
\right.
\end{split}
\centering
\label{Lagrange_r3}
\end{equation*}

\begin{table}[h!]
\centering
\begin{tabular}{c c c c c} 
 \hline
 $x_1(t)$ & $a_k$ & $\omega_k$ & $\phi_k$ & $x_0$\\ [0.5ex]
 \hline  
k=0 & 0.458215129 & 0.044444444 & 2.794701043327 & 0.111456457 \\ 
k=1 & 0.128067684 & 0.088888888 & -0.47929680311 \\
k=2 & 0.057713651 & 0.133333333 & 2.563732159175 \\
k=3 & 0.027083750 & 0.177777777 & -0.75493413322 \\
k=4 & 0.016273630 & 0.222222222 & 2.344062670707 \\[1ex]
 \hline
$y_1(t)$ & $b_k$ & $\alpha_k$ & $\psi_k$ & $y_0$\\ [0.5ex]
\hline 
k=0 & 0.524970110 & 0.044444444 & 1.27664477069 & 0.5811966452 \\
k=1 & 0.143567818 & 0.088888888 & -1.9979199642\\
k=2 & 0.061470599 & 0.133333333 & 1.00439239329\\ 
k=3 & 0.029645570 & 0.177777777 & -2.2605104483\\ 
k=4 & 0.016744548 & 0.222222222 & 0.73885080276\\ [1ex]
 \hline
$x_2(t)$ & $a_k$ & $\omega_k$ & $\phi_k$ & $x_0$\\ [0.5ex]
 \hline  
k=0 & 0.592241079 & 0.044444444 & 1.05406007798 & 0.6152949512 \\ 
k=1 & 0.162118935 & 0.088888888 & -2.2252346931 \\
k=2 & 0.070218733 & 0.133333333 & 0.76737870627 \\
k=3 & 0.033408563 & 0.177777777 & -2.4936543405 \\
k=4 & 0.019369500 & 0.222222222 & 0.49664045558 \\[1ex]
\hline
$y_2(t)$ & $b_k$ & $\alpha_k$ & $\psi_k$ & $y_0$\\ [0.5ex]
\hline 
k=0 & 0.539333087 & 0.044444444 & -0.6283207579 & -0.290158659 \\
k=1 & 0.150404024 & 0.088888888 & 2.38642908793\\
k=2 & 0.066808369 & 0.133333333 & -0.8438848272\\ 
k=3 & 0.031839733 & 0.177777777 & 2.11792385197\\ 
k=4 & 0.018577004 & 0.222222222 & -1.0578825301\\ [1ex]
\hline
$x_3(t)$ & $a_k$ & $\omega_k$ & $\phi_k$ & $x_0$\\ [0.5ex]
 \hline  
k=0 & 0.413994573 & 0.044444444 & -1.1293050516 & -0.391239535 \\ 
k=1 & 0.114373008 & 0.088888888 & 1.88982558405 \\
k=2 & 0.049258308 & 0.133333333 & -1.3459120149 \\
k=3 & 0.024041083 & 0.177777777 & 1.63193352385 \\
k=4 & 0.013347322 & 0.222222222 & -1.5702159732 \\[1ex]
\hline
$y_3(t)$ & $b_k$ & $\alpha_k$ & $\psi_k$ & $y_0$\\ [0.5ex]
\hline 
k=0 & 0.397609775 & 0.044444444 & -2.5603602521 & -0.290818163 \\
k=1 & 0.109689082 & 0.088888888 & 0.43735828509\\
k=2 & 0.048964960 & 0.133333333 & -2.8528515681\\ 
k=3 & 0.022710415 & 0.177777777 & 0.15723662603\\ 
k=4 & 0.013846385 & 0.222222222 & -3.1152974127\\ [1ex]
\hline
\end{tabular}
\caption{The leading 5 terms of parameters of the Lagrange solution.}
\label{table:1}
\end{table}

\section*{The Broucke-Hadjidemetriou-Hénon solution}

The initial conditions of the BHH solution is reported in \textit{Three-body problem — From Newton to supercomputer plus machine learning}\cite{Liao}. The solution is plotted in a rotating reference frame in the original paper. Here, we plot this solution in the fixed center-of-mass reference frame. The masses of three particles are $m_1 = 0.3916, m_2 = 0.8341, m_3 = 1$ and the initial conditions in the center-of-mass reference frame are: \\
$\boldsymbol{r}_1(0) = (-1.3760104789504426,0);$ \\
$\boldsymbol{r}_2(0) = (0.8390195210495575,0);$ \\
$\boldsymbol{r}_3(0) = (-0.16098047895044254,0);$ \\
$\dot{\boldsymbol{r}}_1(0)=(0, -1.00328);$ \\
$\dot{\boldsymbol{r}}_2(0) = (0, -0.53749);$ \\
$\dot{\boldsymbol{r}}_3(0)=(0,0.8412048569999999)$. \\
\begin{figure}[h!]
    \centering
    \includegraphics[width=0.9\textwidth]{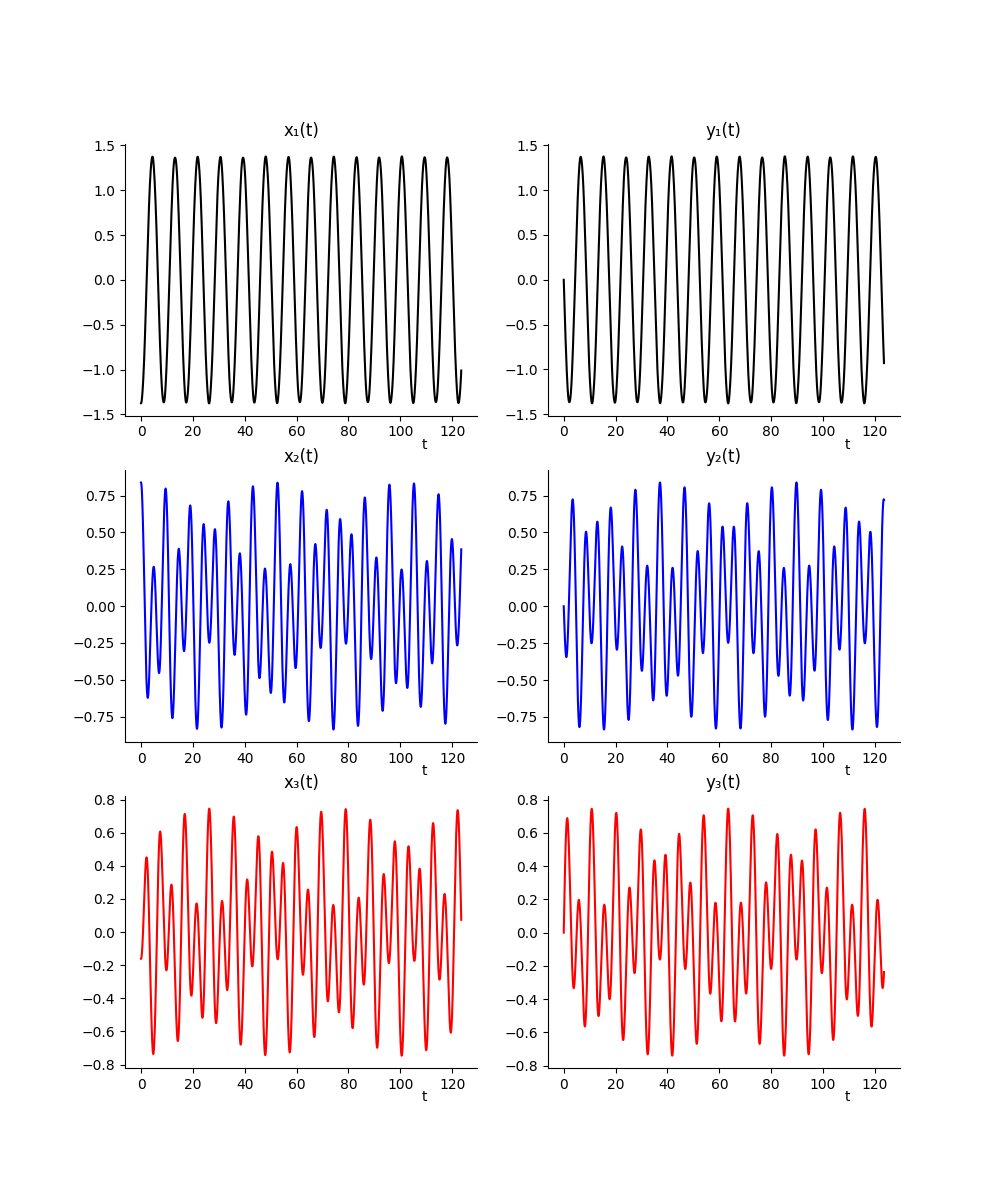} 
    \caption{The $x(t)$ and $y(t)$ of three particles' orbits from numerical solution.}
    \label{BHH_xt_yt}
\end{figure}

It is observed from the numerical solution that the 1st particle's trajectory is close to a circle, and the other two particles' orbits are rose curves. 

\begin{figure}[h!]
    \centering
    \includegraphics[width=0.9\textwidth]{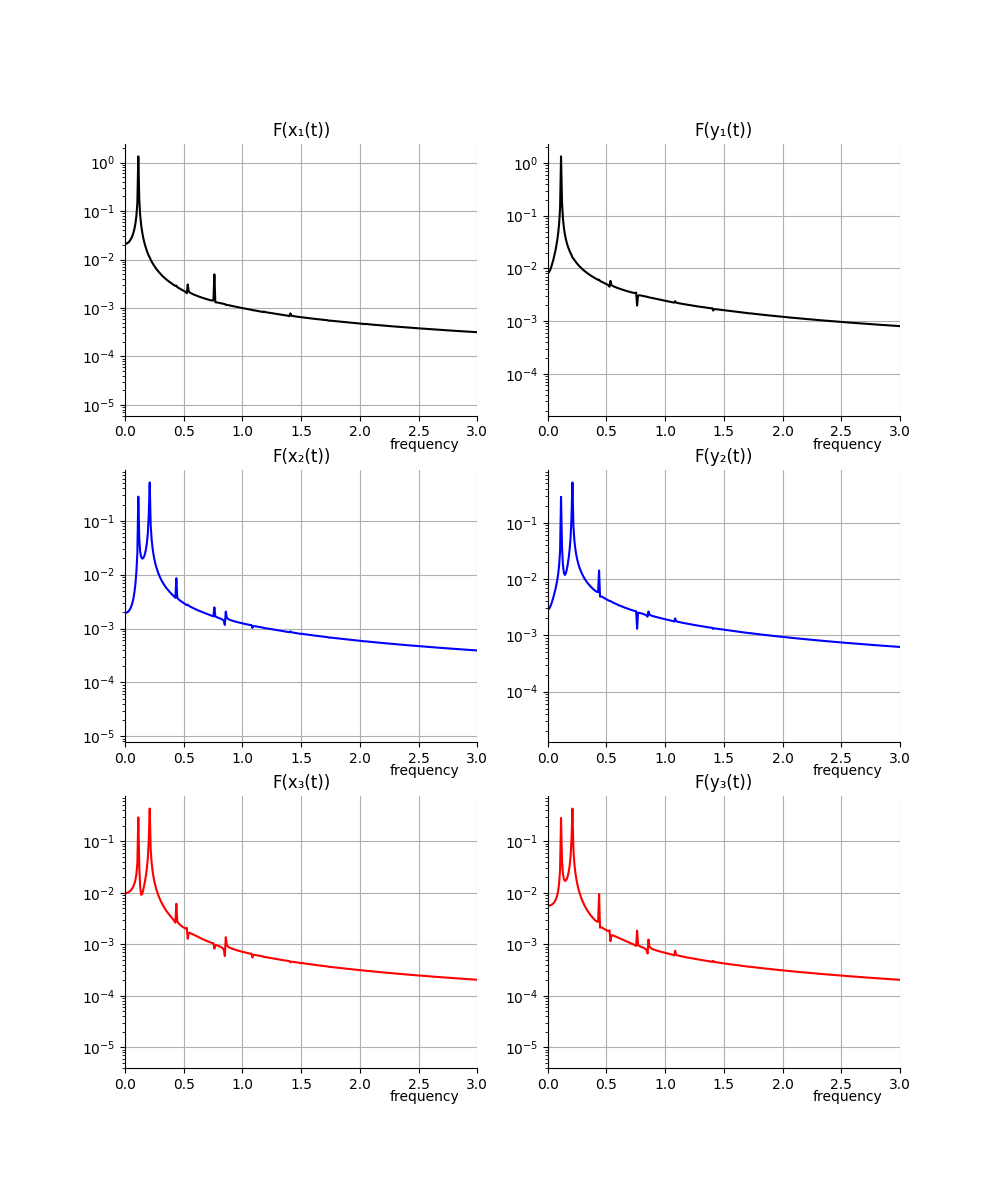} 
    \caption{The Fourier analysis of three particles' orbits.}
    \label{BHH_spectrum}
\end{figure}

For the 1st particle, the dominant frequency 0.113360323 is observed. And for the 2nd and 3rd particles, the two dominant frequencies 0.210526315 and 0.113360323 are observed.

\begin{figure}[h!]
    \centering
    \includegraphics[width=0.9\textwidth]{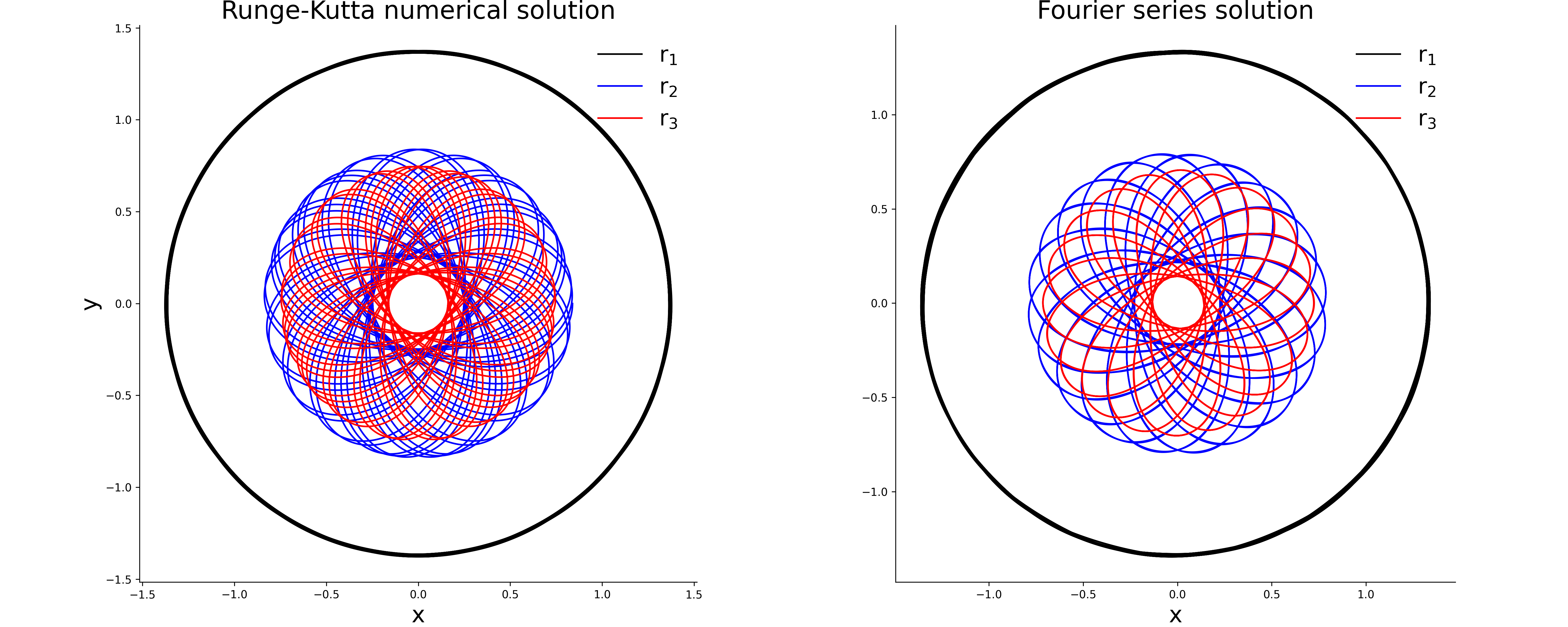} 
    \caption{The Runge-Kutta numerical solution(left) and the Fourier series solution(right) of the BHH solution.}
    \label{BHH_RungeKutta_Fourier}
\end{figure}
Compared with the numerical solution, the Fourier series solution captures main features of this BHH solution.

\begin{equation*}
\begin{split}
\boldsymbol{r}_{1}(t)= \left \{
\begin{array}{ll}
    x_{1}(t) \approx -0.01046621+1.342531089*sin(0.113360323* 2\pi t-1.20107863606) \\
    \hspace{1.5cm} + 0.004937177*sin(0.761133603*2\pi t
    -1.96790832980) + ... \\       
    y_{1}(t) \approx -0.004089034+
    1.334201748*sin(0.113360323*2\pi t-2.76615646368) \\
    \hspace{1.5cm} + 
    0.005794528*sin(0.534412955*2\pi t-0.14167461739) + ... 
\end{array}
\right.
\end{split}
\centering
\label{BHH_r1}
\end{equation*}

\begin{equation*}
\begin{split}
\boldsymbol{r}_{2}(t)= \left \{
\begin{array}{ll}
    x_{2}(t) \approx -0.000977164+0.515918430*sin(0.210526315* 2\pi t+0.941071201543) \\
    \hspace{1.5cm}+ 0.281117986*sin(0.113360323*2\pi t
    +1.964073553194) + ... \\       
    y_{2}(t) \approx -0.001426825+
    0.520344189*sin(0.210526315*2\pi t+2.528423224506) \\
    \hspace{1.5cm}+ 0.288787337*sin(0.113360323*2\pi t+0.348026093027) + ... 
\end{array}
\right.
\end{split}
\centering
\label{BHH_r2}
\end{equation*}

\begin{equation*}
\begin{split}
\boldsymbol{r}_{3}(t)= \left \{
\begin{array}{ll}
    x_{3}(t) \approx 0.0049138117+0.431937467*sin(0.210526315* 2\pi t-2.19121524509) \\
    \hspace{1.5cm}+ 0.291371287*sin(0.113360323*2\pi t+1.921557762968) + ... \\       
    y_{3}(t) \approx 0.0027914638+
    0.429903429*sin(0.210526315*2\pi t-0.62418804791) \\
    \hspace{1.5cm}+ 0.281762576*sin(0.113360323*2\pi t+0.398869623746) + ... 
\end{array}
\right.
\end{split}
\centering
\label{BHH_r3}
\end{equation*}

\begin{table}[h!]
\centering
\begin{tabular}{c c c c c} 
 \hline
 $x_1(t)$ & $a_k$ & $\omega_k$ & $\phi_k$ & $x_0$\\ [0.5ex]
 \hline  
k=0 & 1.342531089 & 0.113360323 & -1.20107863606 & -0.01046621 \\ 
k=1 & 0.004937177 & 0.761133603 & -1.96790832980 \\
k=2 & 0.003054682 & 0.534412955 & -3.14146357710 \\[1ex] 
 \hline
$y_1(t)$ & $b_k$ & $\alpha_k$ & $\psi_k$ & $y_0$\\ [0.5ex]
\hline 
k=0 & 1.334201748 & 0.113360323 & -2.76615646368 & -0.004089034 \\
k=1 & 0.005794528 & 0.534412955 & -0.14167461739\\
k=2 & 0.003079887 & 0.777327935 & 0.060630296230\\ 
k=3 & 0.002385207 & 1.085020242 & 0.014019732641\\ [1ex]
 \hline
$x_2(t)$ & $a_k$ & $\omega_k$ & $\phi_k$ & $x_0$\\ [0.5ex]
 \hline  
k=0 & 0.515918430 & 0.210526315 & 0.941071201543 & -0.000977164 \\ 
k=1 & 0.281117986 & 0.113360323 & 1.964073553194 \\
k=2 & 0.008595243 & 0.437246963 & -1.02607903156 \\
k=3 & 0.002468050 & 0.761133603 & 0.760816665098 \\
k=4 & 0.002061929 & 0.858299595 & 0.291265231939 \\[1ex]
\hline
$y_2(t)$ & $b_k$ & $\alpha_k$ & $\psi_k$ & $y_0$\\ [0.5ex]
\hline 
k=0 & 0.520344189 & 0.210526315 & 2.528423224506 & -0.001426825 \\
k=1 & 0.288787337 & 0.113360323 & 0.348026093027\\
k=2 & 0.014158395 & 0.437246963 & 3.132895866590\\ 
k=3 & 0.002647395 & 0.858299595 & 2.836110044248\\ 
k=4 & 0.002522412 & 0.769230769 & 3.021677521189\\
k=5 & 0.001997198 & 1.085020242 & 3.016855253912\\[1ex]
\hline
$x_3(t)$ & $a_k$ & $\omega_k$ & $\phi_k$ & $x_0$\\ [0.5ex]
 \hline  
k=0 & 0.431937467 & 0.210526315 & -2.19121524509 & 0.0049138117 \\ 
k=1 & 0.291371287 & 0.113360323 & 1.921557762968 \\
k=2 & 0.006126346 & 0.437246963 & 2.048684312211 \\
k=3 & 0.001665549 & 0.550607287 & -2.43852448800 \\
k=4 & 0.001375900 & 0.858299595 & -2.65803699455 \\
k=5 & 0.000970929 & 0.769230769 & -2.61593840674\\[1ex]
\hline
$y_3(t)$ & $b_k$ & $\alpha_k$ & $\psi_k$ & $y_0$\\ [0.5ex]
\hline 
k=0 & 0.429903429 & 0.210526315 & -0.62418804791 & 0.0027914638 \\
k=1 & 0.281762576 & 0.113360323 & 0.398869623746\\
k=2 & 0.009442213 & 0.437246963 & -0.03808809111\\ 
k=3 & 0.001838620 & 0.761133603 & -0.24557958760\\ 
k=4 & 0.001498004 & 0.550607287 & -0.46227147499\\ 
k=5 & 0.001239144 & 0.858299595 & -0.62953500636\\
k=6 & 0.000752096 & 1.085020242 & -0.29563766404\\[1ex]
\hline
\end{tabular}
\caption{The leading 5 terms of parameters of the BHH solution.}
\label{table:1}
\end{table}

\section*{The remarkable figure-8 solution}

Figure-8 solution belongs to the equal-mass-zero-angular-momentum family. The terminology equal-mass-zero-angular-momentum is actually constrains applied to Newton's equations. The equal-mass means the mass of three particles are identical, i.e., $m_1 = m_2 
= m_3 = 1$ and zero-angular-momentum indicates the total angular momentum of the system is zero, i.e., $\sum_1^3 \boldsymbol{r}_i \times m_i \dot{\boldsymbol{r}}_i = 0$. The arbitrary configuration of masses and initial conditions is in ${R^{3+}} \times {R^{12}}$. With the equal-mass-zero-angular-momentum constrain, the choice of one particle's initial condition in ${R^4}$ can determine all particles' initial positions and velocities, therefore, reduce the searching space from ${R^{3+}} \times {R^{12}}$ to $R^4$. 

\begin{figure}[h!]
    \centering
    \includegraphics[width=0.9\textwidth]{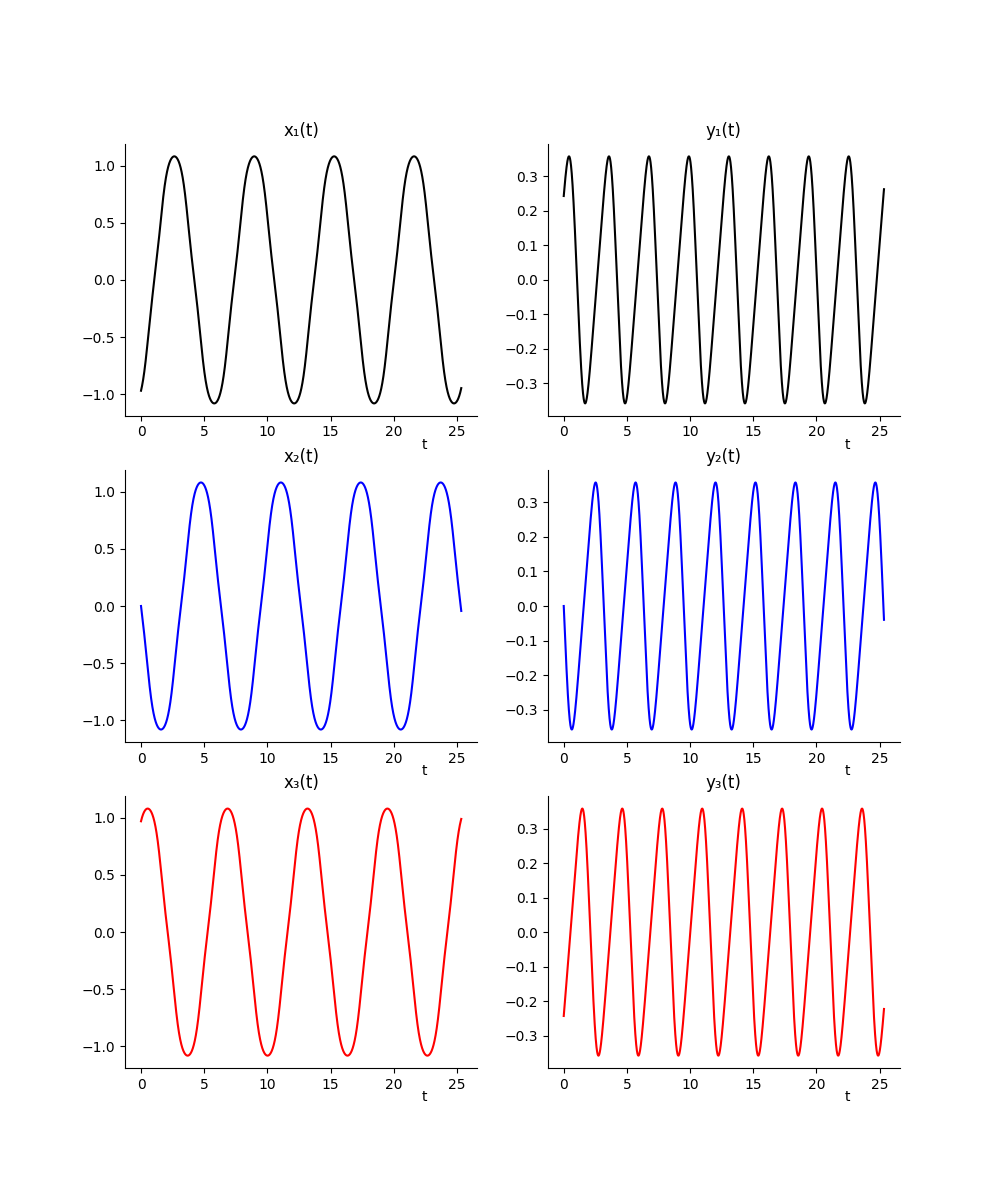} 
    \caption{$x(t)$ and $y(t)$ of the second object that starting from the origin.}
    \label{Figure8_xt_yt}
\end{figure}

Via the Runge-Kutta method, the numerical figure-8 solution is obtained with the initial conditions: \\
$r_1(0) =-r_3(0) = (-0.97000436, 0.24308753);$\\
$r_2(0) = (0,0);$ \\
$\dot{r}_1(0)=\dot{r}_3(0)=(0.4662036850, 0.4323657300);$ \\
$\dot{r}_2(0) = (-0.93240737, -0.86473146)$ \cite{Chenciner}. \\
The $x_i(t)$ and $y_i(t)$ from numerical solution are shown in Figure.\ref{Figure8_xt_yt}. 

\begin{figure}[h!]
    \centering
    \includegraphics[width=0.9\textwidth]{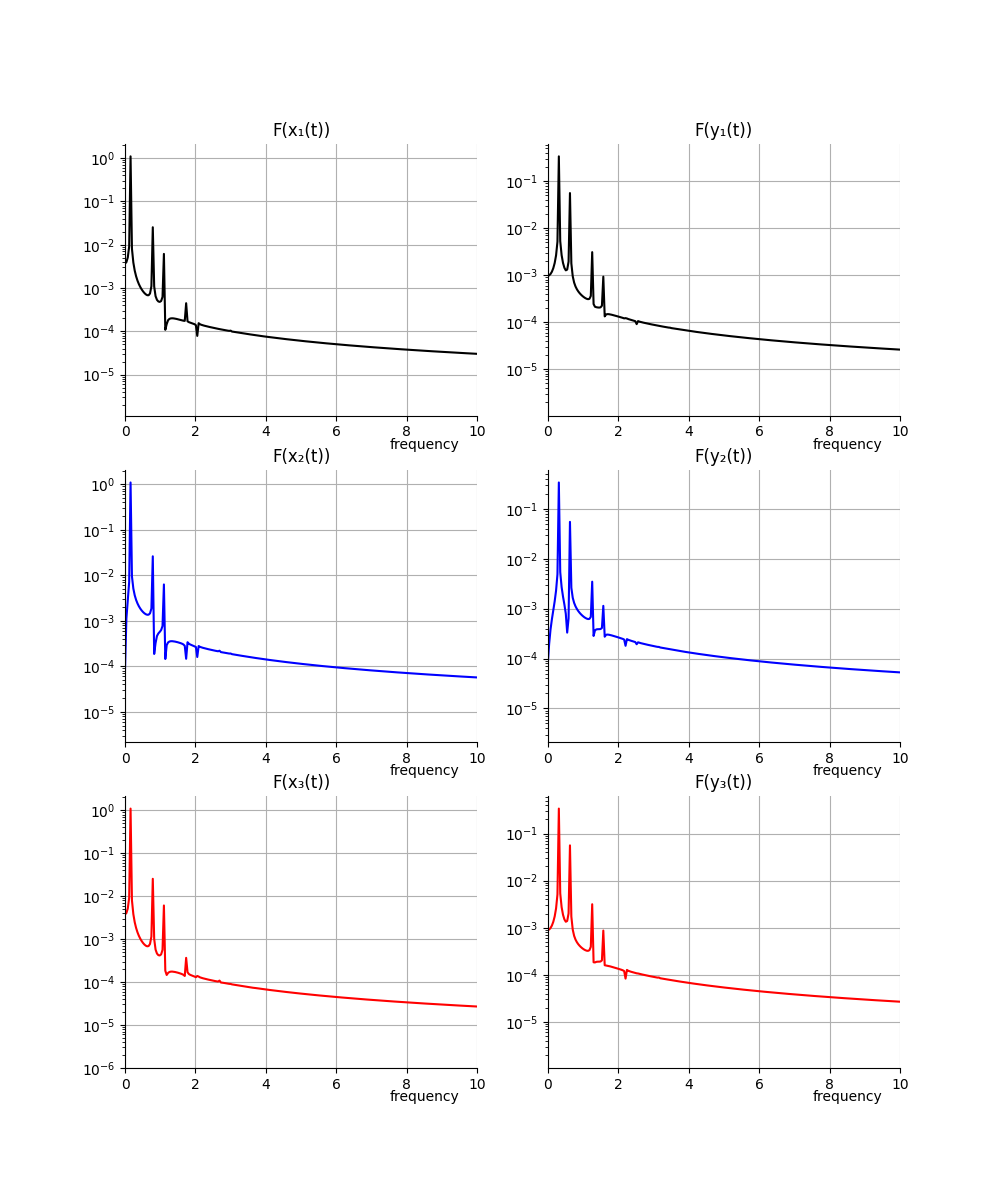}
    \caption{Amplitudes $a_k$ and $b_k$ and frequencies $\omega_k$ and $\alpha_k$ calculated by Fourier transform.}
    \label{Figure8_spectrum}
\end{figure}

\begin{figure}[h!]
    \centering
    \includegraphics[width=0.9\textwidth]{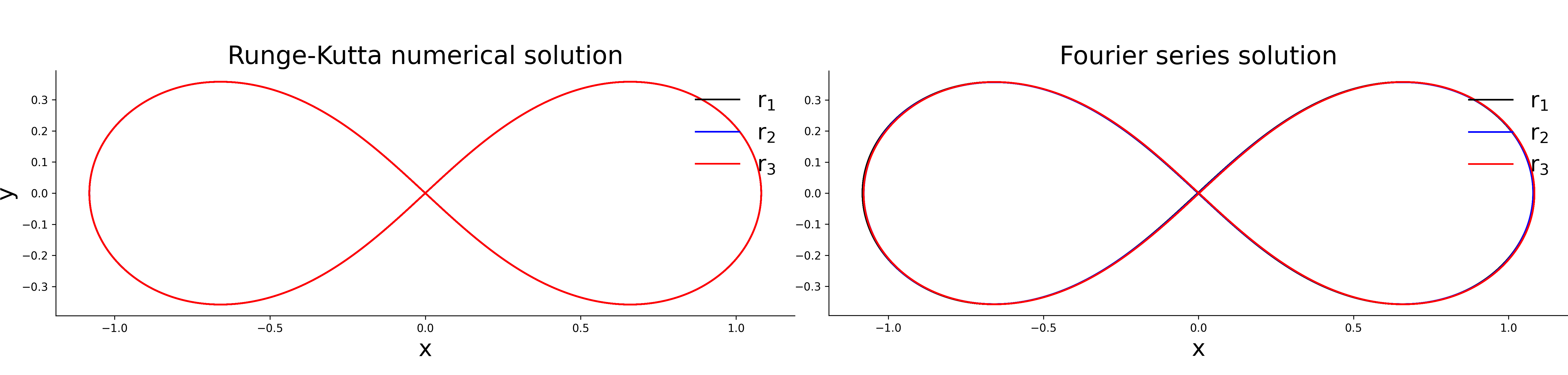}
    \caption{The Runge-Kutta numerical solution(left) and the Fourier series solution(right) of the figure-8 solution.}
    \label{Figure8_RungeKutta_Fourier}
\end{figure}
According to Table.\ref{Figure8_table} of parameters value that the empirical result calculated with above approach, the Fourier series form of the figure-8 solution read as:
\begin{equation*}
\begin{split}
\boldsymbol{r}_{1}(t)= \left \{
\begin{array}{ll}
    x_1(t) \approx -0.00175270094 +1.101277971*sin(0.1577909* 2\pi t-1.02503404) \\
    \hspace{1.5cm}+ 0.025369562*sin(0.7889546*2\pi t-1.99794905) + ... \\       
    y_1(t) \approx 0.000462195787 + 0.338688707*sin(0.3155818*2\pi t+1.093460426) \\
    \hspace{1.5cm}+ 0.055990686*sin(0.6311637*2\pi t-0.9655015) + ... 
\end{array}
\right.
\end{split}
\centering
\label{Figure8_r1}
\end{equation*}

\begin{equation*}
\begin{split}
\boldsymbol{r}_{2}(t)= \left \{
\begin{array}{ll}
    x_2(t) \approx 1.09965496*sin(0.1577909* 2\pi t-3.11854626) \\
    \hspace{1.5cm}+ 0.026214418*sin(0.7889546*2\pi t+0.111859893) + ... \\       
    y_2(t) \approx 338493613*sin(0.3155818*2\pi t-3.09542668) \\
    \hspace{1.5cm}+ 0.054948979*sin(0.6311637*2\pi t-3.04836475) + ... 
\end{array}
\right.
\end{split}
\centering
\label{Figure8_r2}
\end{equation*}

\begin{equation*}
\begin{split}
\boldsymbol{r}_{3}(t)= \left \{
\begin{array}{ll}
    x_3(t) \approx 0.001792361092+1.101350189*sin(0.1577909* 2\pi t+1.071018304) \\
    \hspace{1.5cm}+ 0.025440532*sin(0.7889546*2\pi t+2.226105089) + ... \\       
    y_3(t) \approx -0.00042626081+0.3386211*sin(0.3155818*2\pi t-1.00145851) \\
    \hspace{1.5cm}+ 0.056057362*sin(0.6311637*2\pi t+1.149944682) + ... 
\end{array}
\right.
\end{split}
\centering
\label{Figure8_r3}
\end{equation*}

Based on the frequencies obtained from Fourier transform spectrum, the period T of the figure-8 solution is about $1/0.1577909 \approx 6.3375$. The 1st and 3rd particle's trajectories can be obtained by shifting $r_2(t)$ $\frac{T}{3}$ forward and backward, i.e., $\boldsymbol{r}_1(t) = \boldsymbol{r}_2(t-\frac{T}{3})$ and $\boldsymbol{r}_3(t) = \boldsymbol{r}_2(t+\frac{T}{3})$. 

 
\begin{table}[h!]
\centering
\begin{tabular}{c c c c c} 
\hline
$x_1(t)$ & $a_k$ & $\omega_k$ & $\phi_k$ & $x_0$\\ [0.5ex]
\hline  
k=0 & 1.101277971 & 0.1577909 & -1.02503404 & -0.00175270094 \\ 
k=1 & 0.025369562 & 0.7889546 & -1.99794905 \\
k=2 & 0.006149148 & 1.1045364 & 2.269477627 \\
k=3 & 0.000446775 & 1.7357001 & 1.669351226 \\[1ex]
\hline
$y_1(t)$ & $b_k$ & $\alpha_k$ & $\psi_k$ & $y_0$\\ [0.5ex]
\hline 
k=0 & 0.338688707 & 0.3155818 & 1.093460426 & 0.000462195787 \\
k=1 & 0.055990686 & 0.6311637 & -0.96550150\\
k=2 & 0.003089890 & 1.2623274 & -1.98038230\\ 
k=3 & 0.000929775 & 1.5779092 & 2.452160278\\ [1ex]
\hline
$x_2(t)$ & $a_k$ & $\omega_k$ & $\phi_k$ & $x_0$\\ [0.5ex]
\hline  
k=0 & 1.099654960 & 0.1577909 & -3.11854626 & -3.96704690E-5 \\ 
k=1 & 0.026214418 & 0.7889546 & 0.111859893 \\
k=2 & 0.006308942 & 1.1045364 & 0.148305341 \\
k=3 & 0.000357050 & 1.3412228 & -0.03773205 \\[1ex]
\hline
$y_2(t)$ & $b_k$ & $\alpha_k$ & $\psi_k$ & $y_0$\\ [0.5ex]
\hline 
k=0 & 0.338493613 & 0.3155818 & -3.09542668 & -3.59367564E-5 \\
k=1 & 0.054948979 & 0.6311637 & -3.04836475\\
k=2 & 0.003473734 & 1.2623274 & 0.162702926\\ 
k=3 & 0.001138562 & 1.5779092 & 0.165080684\\ [1ex]
\hline
$x_3(t)$ & $a_k$ & $\omega_k$ & $\phi_k$ & $x_0$\\ [0.5ex]
\hline  
k=0 & 1.101350189 & 0.1577909 & 1.071018304 & 0.001792361092 \\ 
k=1 & 0.025440532 & 0.7889546 & 2.226105089 \\
k=2 & 0.006078589 & 1.1045364 & -1.95454416 \\
k=3 & 0.000365579 & 1.7357001 & -1.13303700 \\[1ex]
\hline
$y_3(t)$ & $b_k$ & $\alpha_k$ & $\psi_k$ & $y_0$\\ [0.5ex]
\hline 
k=0 & 0.338621100 & 0.3155818 & -1.00145851 & -0.00042626081 \\
k=1 & 0.056057362 & 0.6311637 & 1.149944682\\
k=2 & 0.003153474 & 1.2623274 & 2.335082741\\ 
k=3 & 0.000870805 & 1.5779092 & -2.04354903\\ [1ex]
\hline
\end{tabular}
\caption{The leading order parameters of the figure-8 solution.}
\label{Figure8_table}
\end{table}

\section*{The IA100 solution}
The IA100 solution is another example of the equal-mass-zero-angular-momentum family with a period of 163.9101889958 and its initial conditions\cite{Li} are: \\
$\boldsymbol{r}_1(0) = (0,0);$ \\
$\boldsymbol{r}_2(0) =-\boldsymbol{r}_3(0) = (-1, 0);$ 
$\dot{\boldsymbol{r}}_1(0) = (-0.1341521554, -1.1779255784)$;\\
$\dot{\boldsymbol{r}}_2(0)=\dot{\boldsymbol{r}}_3(0)=(0.0670760777, 0.5889627892)$. \\

\begin{figure}[h!]
    \centering
    \includegraphics[width=0.9\textwidth]{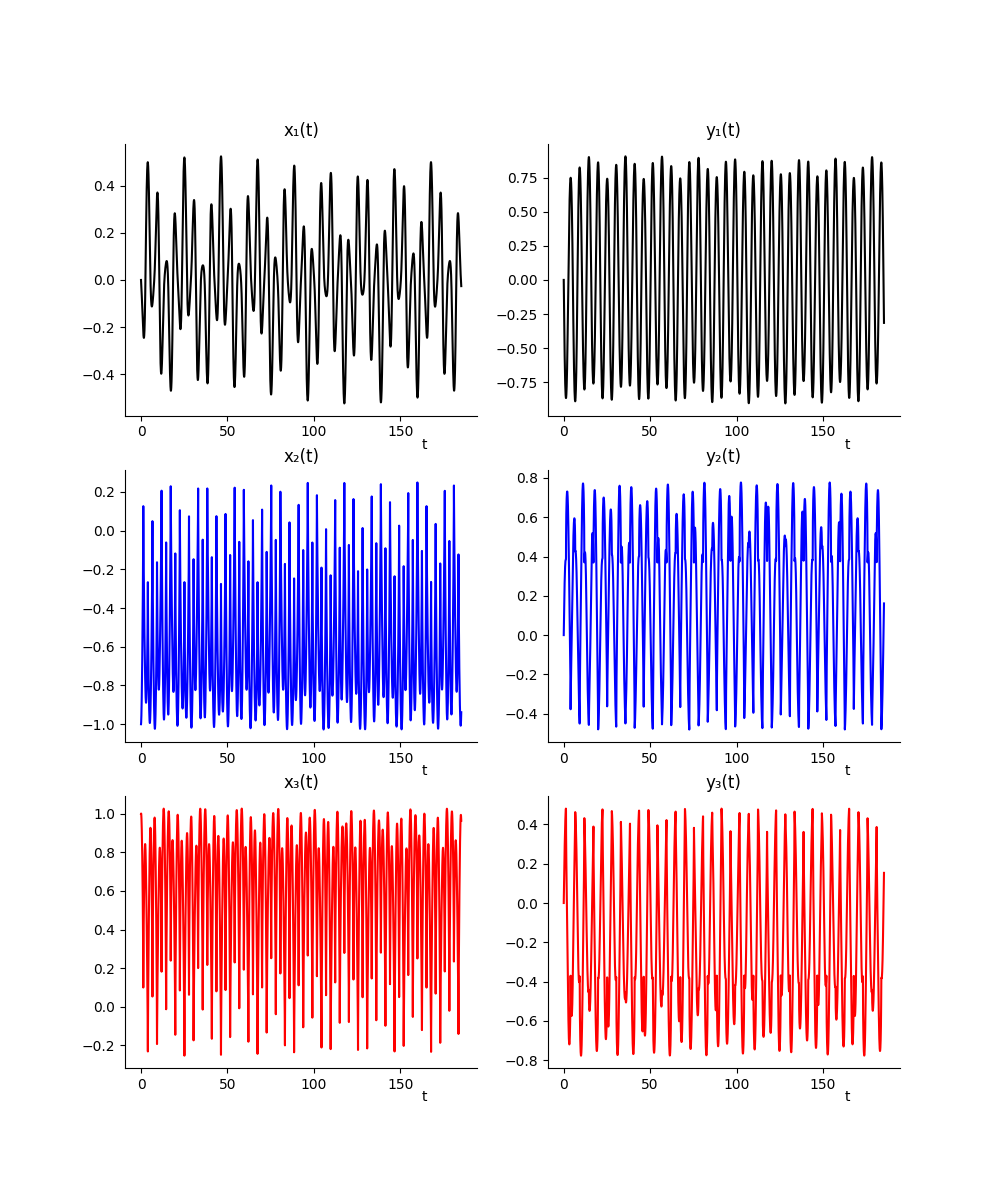} 
    \caption{The $x(t)$ and $y(t)$ of three particles' orbits from numerical solution.}
    \label{IA100_xt_yt}
\end{figure}

\begin{figure}[h!]
    \centering
    \includegraphics[width=0.9\textwidth]{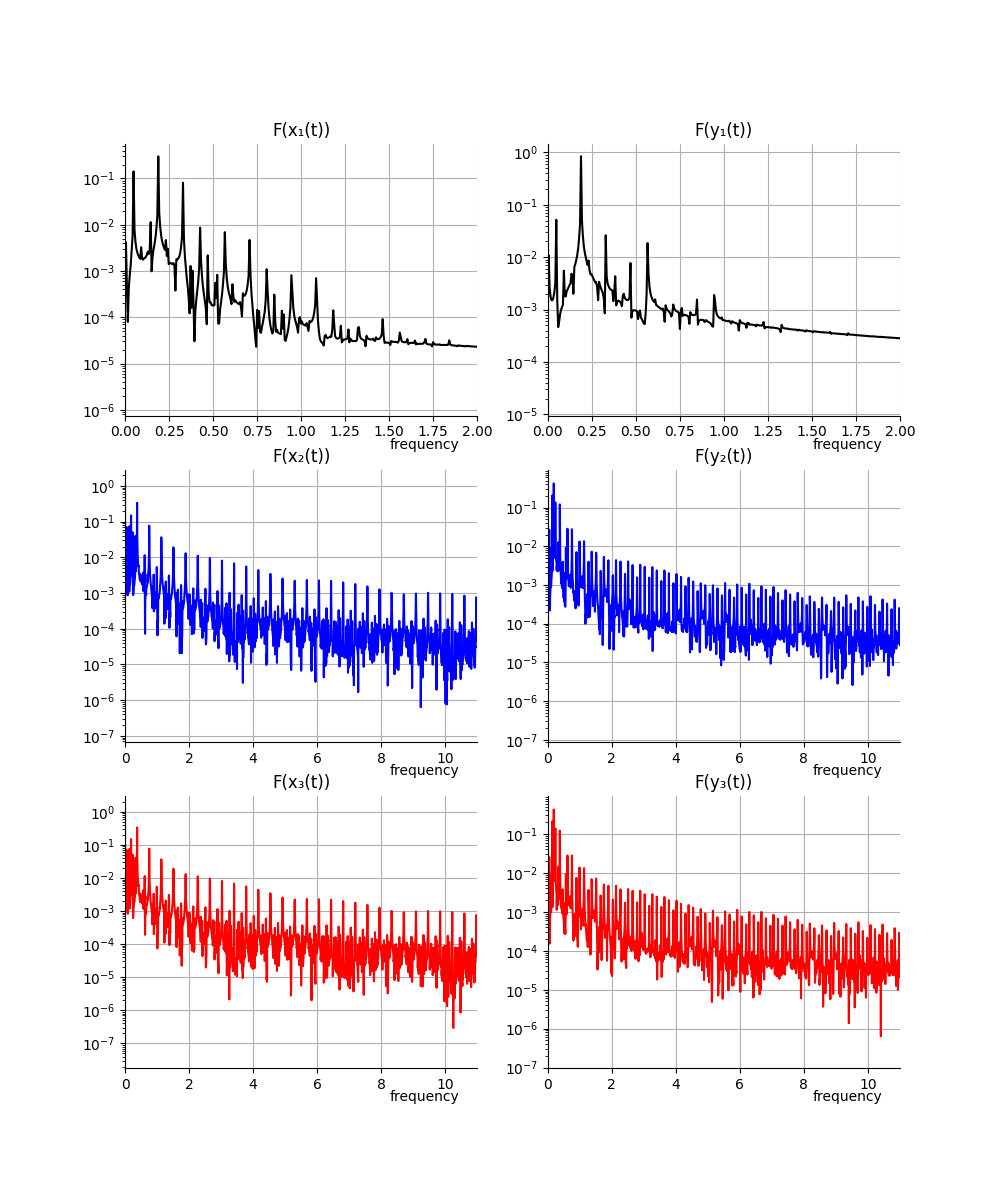} 
    \caption{The Fourier analysis of three particles' orbits.}
    \label{IA100_spectrum}
\end{figure}
The 1st particle's spectrum is narrower than the other two particles' spectrum. And the 2nd and 3rd particles' motion is symmetrical with the identical spectrum. 

\begin{figure}[h!]
    \centering
    \includegraphics[width=0.9\textwidth]{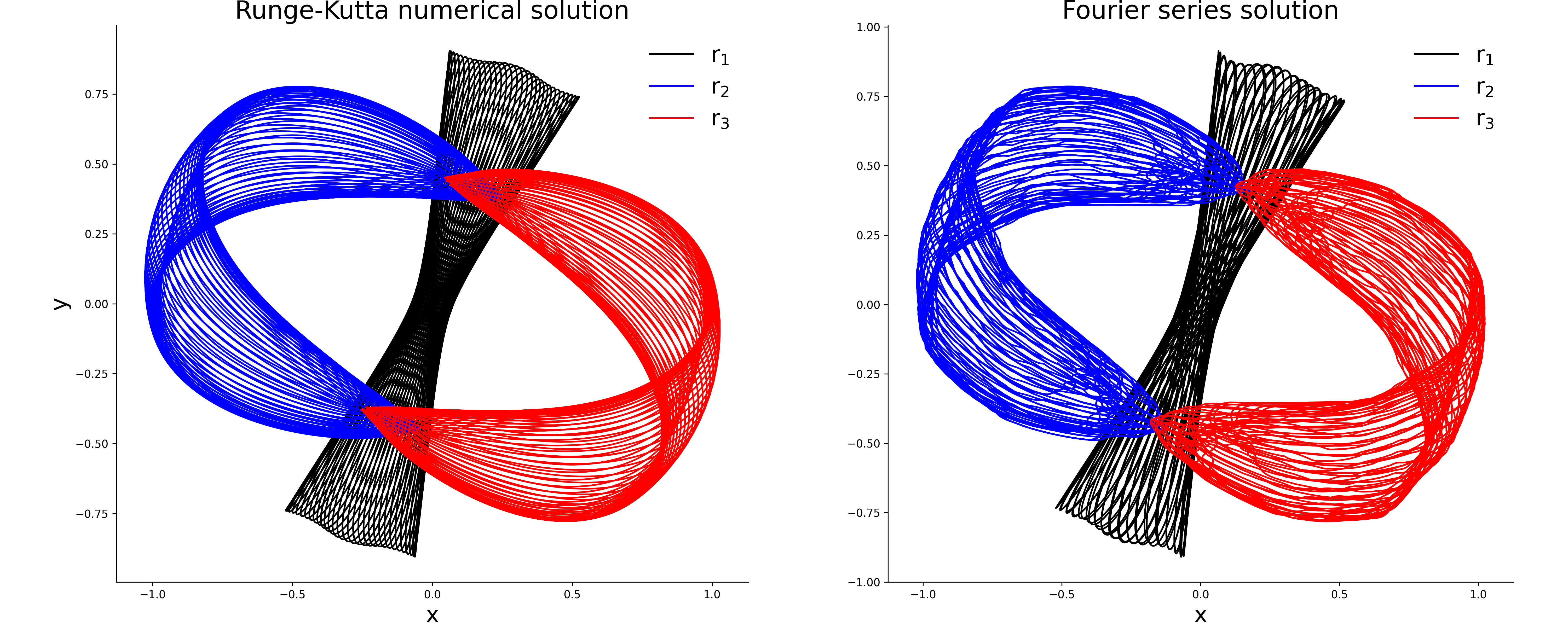} 
    \caption{The Runge-Kutta numerical solution(left) and the Fourier series solution(right) of IA100 with initial conditions $r_2(0) =-r_3(0) = (-1, 0);$ $r_1(0) = (0,0);$ $\dot{r}_2(0)=\dot{r}_3(0)=(0.0670760777, 0.5889627892);$ $\dot{r}_1(0) = (-0.1341521554, -1.1779255784)$.}
    \label{IA100_RungeKutta_Fourier}
\end{figure}

\begin{equation*}
\begin{split}
\boldsymbol{r}_{1}(t)= \left \{
\begin{array}{ll}
    x_{1}(t) \approx 0.000252558+0.300828564*sin(0.188933873* 2\pi t-2.99207959848) \\
    \hspace{1.5cm} + 0.142242548*sin(0.048582995*2\pi t +0.140413868368) + ... \\       
    y_{1}(t) \approx 0.0002810119+
    0.842954275*sin(0.188933873*2\pi t-2.99165536036) \\
    \hspace{1.5cm} + 
    0.052005297*sin(0.048582995*2\pi t-2.99309613061) + ... 
\end{array}
\right.
\end{split}
\centering
\label{IA100_r1}
\end{equation*}

\begin{equation*}
\begin{split}
\boldsymbol{r}_{2}(t)= \left \{
\begin{array}{ll}
    x_{2}(t) \approx -0.6268658298+0.336750679*sin(0.377867746* 2\pi t-1.2724363714) \\
    \hspace{1.5cm}+ 0.150568677*sin(0.188933873*2\pi t+0.13536448564) + ... \\       
    y_{2}(t) \approx 0.22299028818+
    0.421648679*sin(0.188933873*2\pi t+0.14811415819) \\
    \hspace{1.5cm}+ 0.207468245*sin(0.140350877*2\pi t-1.5515273324) + ... 
\end{array}
\right.
\end{split}
\centering
\label{IA100_r2}
\end{equation*}

\begin{equation*}
\begin{split}
\boldsymbol{r}_{3}(t)= \left \{
\begin{array}{ll}
    x_{3}(t) \approx 0.62661323732+0.336640447*sin(0.377867746* 2\pi t+1.86950221549) \\
    \hspace{1.5cm}+ 0.150291116*sin(0.188933873*2\pi t+0.16368567521) + ... \\       
    y_{3}(t) \approx -0.2232716772+
    0.421306985*sin(0.188933873*2\pi t+0.15176026845) \\
    \hspace{1.5cm}+ 0.208306433*sin(0.140350877*2\pi t+1.57102414710) + ... 
\end{array}
\right.
\end{split}
\centering
\label{IA100_r3}
\end{equation*}

\begin{table}[h!]
\centering
\begin{tabular}{c c c c c} 
 \hline
 $x_1(t)$ & $a_k$ & $\omega_k$ & $\phi_k$ & $x_0$\\ [0.5ex]
 \hline  
k=0 & 0.300828564 & 0.188933873 & -2.99207959848 & 0.000252558 \\ 
k=1 & 0.142242548 & 0.048582995 & 0.140413868368 \\
k=2 & 0.081019946 & 0.329284750 & 0.159397289623 \\
k=3 & 0.011378304 & 0.145748987 & -2.74934949853 \\
k=4 & 0.008647818 & 0.426450742 & 0.434762694467 \\[1ex]
 \hline
$y_1(t)$ & $b_k$ & $\alpha_k$ & $\psi_k$ & $y_0$\\ [0.5ex]
\hline 
k=0 & 0.842954275 & 0.188933873 & -2.99165536036 & 0.0002810119 \\
k=1 & 0.052005297 & 0.048582995 & -2.99309613061\\
k=2 & 0.026320194 & 0.329284750 & -2.97710641215\\ 
k=3 & 0.018617612 & 0.566801619 & -2.67160002092\\ 
k=4 & 0.010883693 & 0.005398110 & 0.426682145280\\ [1ex]
 \hline
$x_2(t)$ & $a_k$ & $\omega_k$ & $\phi_k$ & $x_0$\\ [0.5ex]
 \hline  
k=0 & 0.336750679 & 0.377867746 & -1.2724363714 & -0.6268658298 \\ 
k=1 & 0.150568677 & 0.188933873 & 0.13536448564 \\
k=2 & 0.077123310 & 0.755735492 & 2.16632993021 \\
k=3 & 0.075091249 & 0.140350877 & -1.5401474926 \\
k=4 & 0.071026096 & 0.048582995 & -2.9923468552 \\[1ex]
\hline
$y_2(t)$ & $b_k$ & $\alpha_k$ & $\psi_k$ & $y_0$\\ [0.5ex]
\hline 
k=0 & 0.421648679 & 0.188933873 & 0.14811415819 & 0.22299028818 \\
k=1 & 0.207468245 & 0.140350877 & -1.5515273324\\
k=2 & 0.136652390 & 0.237516869 & -1.3008245351\\ 
k=3 & 0.120815570 & 0.377867746 & 1.87892090710\\ 
k=4 & 0.028376874 & 0.615384615 & 2.17745126561\\ [1ex]
\hline
$x_3(t)$ & $a_k$ & $\omega_k$ & $\phi_k$ & $x_0$\\ [0.5ex]
 \hline  
k=0 & 0.336640447 & 0.377867746 & 1.86950221549 & 0.62661323732 \\ 
k=1 & 0.150291116 & 0.188933873 & 0.16368567521 \\
k=2 & 0.077103772 & 0.755735492 & -0.9759538438 \\
k=3 & 0.076073627 & 0.140350877 & 1.56683696919 \\
k=4 & 0.071222682 & 0.048582995 & -3.0099840785 \\[1ex]
\hline
$y_3(t)$ & $b_k$ & $\alpha_k$ & $\psi_k$ & $y_0$\\ [0.5ex]
\hline 
k=0 & 0.421306985 & 0.188933873 & 0.15176026845 & -0.2232716772 \\
k=1 & 0.208306433 & 0.140350877 & 1.57102414710\\
k=2 & 0.137680982 & 0.237516869 & 1.87966418328\\ 
k=3 & 0.120606519 & 0.377867746 & -1.2808151613\\ 
k=4 & 0.027997398 & 0.755735492 & 2.18161656262\\ [1ex]
\hline
\end{tabular}
\caption{The leading 5 terms of parameters of the IA100 solution.}
\label{table:1}
\end{table}

\section*{Conclusion}

To overcome the difficulties in describing and solving curves of Newton's equations, the Antikythera algorithm is introduced and four examples from the Lagrange, BHH, equal-mass-zero-angular-momentum are presented along with their spectrum. This technique helps us calculate the Fourier series solution of three-body problem and creates good conditions for the further study and analysis of problems, such as Saari's conjecture\cite{Saari} and Smale's sixth problem\cite{Smale}. 

We look forward to exploring the application of similar technique in astrophysics and express the celestial orbits in the form of Fourier series, such as planets' orbits in solar system, the triple stars system Alpha Centauri and so forth. In addition, due to the similarity between Newton's gravitational force and Coulombian force, insights into the Newtonian n-body problem can be easily transferred to Coulombian n-body problem and lead to a better understanding of the solid state, conductivity, and phase changing of matter. 

From the comparison between Runge-Kutta numerical solution and Fourier series solution, the discrepancies are mainly coming from the measurement error from Fourier analysis and only finite terms can be taken into computer's calculation. It is worthwhile noting that the goal is not purely to fit the numerical solution but to solve Newton's equations. Thus, all calculated curves shall be plugged back in Newton's equations and prove they are satisfied, which involves the challenge of computing operations between very lengthy Fourier series. 

Furthermore, the Fourier analysis spectrum from the experiments reveals that the spectrum has very fine structure and deep math discipline underhood. The spectrum theorem and Floquet theory might shed light on this phenomenon and help us analytically determine the general terms in the Fourier series, which will bring us to the new level understanding of this problem.

\section*{Acknowledgments}
This work is partly supported by the Science and Technology Research Project of Jiangxi Provincial Department of Education under Grant no.GJJ2200377 and the National Natural Science Foundation of China under Grant no.62261028.



\end{document}